# Software-Defined Networking: State of the Art and Research Challenges

Manar Jammal[a1], Taranpreet Singh[a], Abdallah Shami[a], RasoolAsal[b], and Yiming Li[c]
[a] Department of Electrical and Computer Engineering, Western University, Canada
[b] British Telecom, UK
[c] StarTech.com, Canada

*Abstract*—Plug-and-play information technology (IT) infrastructure has been expanding very rapidly in recent years. With the advent of cloud computing, many ecosystem and business paradigms are encountering potential changes and may be able to eliminate their IT infrastructure maintenance processes. Real-time performance and high availability requirements have induced telecom networks to adopt the new concepts of the cloud model: software-defined networking (SDN) and network function virtualization (NFV). NFV introduces and deploys new network functions in an open and standardized IT environment, while SDN aims to transform the way networks function. SDN and NFV are complementary technologies; they do not depend on each other. However, both concepts can be merged and have the potential to mitigate the challenges of legacy networks. In this paper, our aim is to describe the benefits of using SDN in a multitude of environments such as in data centers, data center networks, and Network as Service offerings. We also present the various challenges facing SDN, from scalability to reliability and security concerns, and discuss existing solutions to these challenges.

## 1. INTRODUCTION

Today's Internet applications require the underlying networks to be fast, carry large amounts of traffic, and to deploy a number of distinct, dynamic applications and services. Adoption of the concepts of "inter-connected data centers" and "server virtualization" has increased network demand tremendously. In addition to various proprietary network hardware, distributed protocols, and software components, legacy networks are inundated with switching devices that decide on the route taken by each packet individually; moreover, the data paths and the decision-making processes for switching or routing are collocated on the same device. This situation is elucidated in Fig. 1. The decision-making capability or network intelligence is distributed across the various network hardware components. This makes the introduction of any new network device or service a tedious job because it requires reconfiguration of each of the numerous network nodes.

Legacy networks have become difficult to automate [1, 2].Networks today depend on IP addresses to identify and locate servers and applications. This approach works fine for static networks where each physical device is recognizable by an IP address, but is extremely laborious for large virtual networks. Managing such complex environments using traditional networks is time-consuming and expensive, especially in the case of virtual machine (VM) migration and network configuration. To simplify the task of managing large virtualized networks, administrators must resolve the physical infrastructure concerns that increase management complexity. In addition, most modern-day vendors use control-plane software to optimize data flow to achieve high performance and competitive advantage [2]. This switch-based control-plane paradigm gives network administrators very little opportunity to increase data-flow efficiency across the network as a whole. The rigid structure of legacy networks prohibits programmability to meet the variety of client requirements, sometimes forcing vendors into deploying complex and fragile programmable management systems. In addition, vast teams of network administrators are employed to make thousands of changes manually to network components [2, 3].

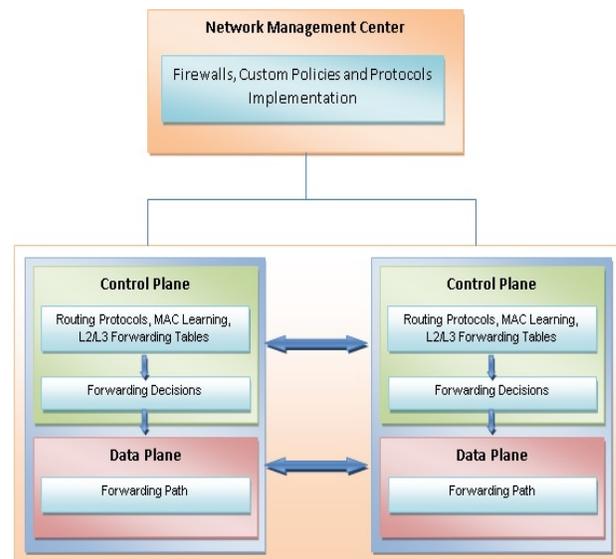

Figure 1: Inflexible Legacy Infrastructure

The demand for services and network usage is growing rapidly. Although growth drivers such as video traffic, big data, and mobile usage augment revenues, they pose significant challenges for network operators [4]. Mobile and Telco operators are encountering spectrum congestion, the shift to internet protocol (IP), and increased mobile users. Concurrently, data-center operators are facing tremendous growth in the number of servers and virtual machines, increasing server-to-server communication traffic. In order to tackle these challenges, operators require a network that is efficient, flexible, agile, and scalable.

Inspired by the words of Marc Andreesen, "software is eating the world", software-defined networking (SDN) and virtualization are poised to be the solutions that overcome the challenges described above. SDN operates on an aggregated and centralized control plane that might be a promising solution for network management and control problems. The main idea behind SDN is to separate the forwarding/data plane from the control plane while providing programmability on the control plane, as illustrated in Fig.2.

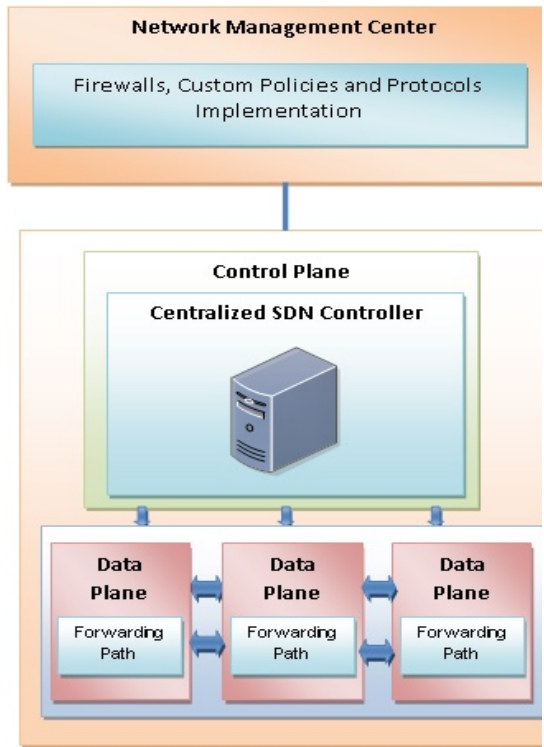

**Figure 2: SDN Architecture**

Despite its obvious advantages and its ability to simplify networks, SDN encounters some technical challenges that can restrict its functionality and performance in cloud computing, information technology (IT) organizations, and networking enterprises. Compared to recent surveys [5, 6], this paper tackles most of the SDN challenges with their causes and existing solutions in a comprehensive and detailed manner. It addresses reliability, scalability, latency, controller placement, recent hardware shortages, and security issues. Overcoming these challenges might assist IT organizations and network enterprises in exploring and improving the various opportunities and functionalities of SDN.

In this paper, Section II defines SDN and discusses its architecture and its protocol, OpenFlow. The concept of network virtualization (NV) is elucidated in Section III, with a discussion of how NV has emerged as a potential solution to the current ossified network architecture and offers benefits that can rapidly alter both the networking and cloud-computing industries. Section IV discusses various SDN applications in data-center networks and Network as a Service. Section V analyzes the various challenges facing SDN, their causes, and their recent solutions. Finally, the last section summarizes various research initiatives in the SDN field, starting from SDN prototypes, development tools and languages, and virtualization implementations using SDN and ending with the various SDN vendors.

## 2. SOFTWARE-DEFINED NETWORKING AND OPENFLOW ARCHITECTURE

Most current network devices have control and data-flow functionalities operating on the same device. The only control available to a network administrator is from the network management plane, which is used to configure each network node separately. The static nature of current network devices does not permit detailed control-plane configuration. This is exactly where software-defined networking comes into the picture. The ultimate goal of SDN as defined in [7] is to "provide open user-controlled management of the forwarding hardware of a network element." SDN operates on the idea of centralizing control-plane intelligence, but keeping the data plane separate. Thus, the network hardware devices keep their switching fabric (data plane), but hand over their intelligence (switching and routing functionalities) to the controller. This enables the administrator to configure the network hardware directly from the controller. This centralized control of the entire network makes the network highly flexible [8, 9].

### 2.1 SDN Architecture

Compared to legacy networks, there are four additional components in SDN [8, 9, and 10].

*1) Control Plane*
The control plane/controller presents an abstract view of the complete network infrastructure, enabling the administrator to apply custom policies/protocols across the network hardware. The network operating system (NOX) controller is the most widely deployed controller.

*2) Northbound Application Interfaces*
The "northbound" application programming interfaces (APIs) represent the software interfaces between the software modules of the controller platform and the SDN applications running atop the network platform. These APIs expose universal network abstraction data models and functionality for use by network applications. The "northbound APIs" are open source-based.

*3) East-West Protocols*
In the case of a multi-controller-based architecture, the East-West interface protocol manages interactions between the various controllers.

*4) Data Plane and Southbound Protocols*
The data plane represents the forwarding hardware in the SDN network architecture. Because the controller needs to communicate with the network infrastructure, it requires certain protocols to control and manage the interface between various pieces of network equipment. The most popular "southbound protocol" is the OpenFlow protocol. The following section explains OpenFlow and its architecture.

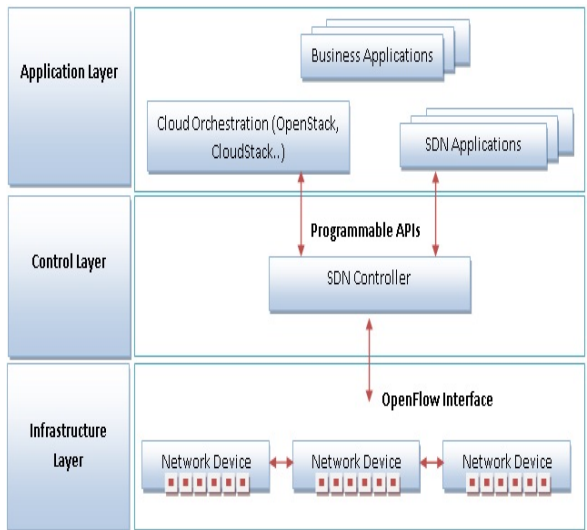

Figure 3: Basic SDN-Based Network Architecture

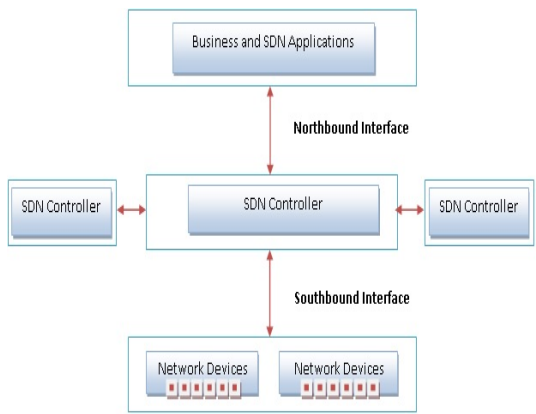

Figure 4: API Directionality in SDN Architecture

## 2.2 SDN Benefits

SDN provides several benefits to address the challenges facing legacy network architectures.

*1) Programmability of the Network:* By implementing a new orchestration level, SDN can tackle the inflexibility and complexity of the traditional network. SDN provides enterprises with the ability to control their networks programmatically and to scale them without affecting performance, reliability, or the user experience [4]. The data- and control-plane abstractions constitute the immense worth of SDN. By eliminating the complexity of the infrastructure layer and adding visibility for applications and services, SDN simplifies network management and brings virtualization to the network. It abstracts flow control from individual devices to the network level. Network-wide data-flow control gives administrators the power to define network flows that meet connectivity requirements and address the specific needs of discrete user communities.

With the SDN approach, network administrators no longer need to implement custom policies and protocols on each device in the network separately. In the general SDN architecture, control-plane functions are separated from physical devices and are performed by an external controller (e.g., standard server running SDN software). SDN provides programmability on the control plane itself, through which changes can be implemented and disseminated either to a specific device or throughout the network hardware on a secure channel. This approach promises to facilitate the integration of new devices into the existing architecture. The SDN controller improves the traffic engineering capabilities of the network operators using video traffic. It enables network operators to control their congestion hot spots and reduces the complexity of traffic engineering [4].

*2) The Rise of Virtualization*: SDN is a promising opportunity for managing hyper-scale data centers (DCs). Data centers raise significant scalability issues, especially with the growth of virtual machines (VMs) and their migration. Moving a VM and updating the media access control (MAC) address table using traditional network architecture may interrupt the user experience and applications.

Therefore, network virtualization, which can be seen as an SDN application, offers a prominent opportunity for hyper-scale data centers. It provides tunnels that can abstract the MAC address from the infrastructure layer, enabling Layer 2 traffic to run over Layer 3 overlays and simplifying VM deployment and migration in the network [4].

Furthermore, SDN enables multi-tenant hosting providers to link their physical and virtual servers, local and remote facilities, and public and private clouds into a single logical network. As a result, each customer will have an isolated view of the network provider. SDN adds a virtualization layer to the fabric architecture of the cloud providers. This enables their tenants to obtain various views over the data-center network (DCN) according to their demands.

SDN is a promising approach for offering Networks as a Service (NaaS) which will enable flexible service models and virtual network operators and endow enterprises with the ability to control DCs and their traffic. This paper introduces the benefits of NaaS and its consolidation with SDN using different cloud models.

*3) Device Configuration and Troubleshooting:* With SDN, device configuration and troubleshooting can be done from a single point on the network which pushes us closer to realizing the ultimate goal of "a dynamic network" that can be configured and made adaptable according to needs. SDN also provides the capability to encourage innovation in the networking field by offering a programmable platform for experiments on novel protocols and policies using production traffic. Separating data flows from test flows facilitates the adoption of newer protocols and ideas into the networking domain [2, 3].

From a broader perspective, SDN offers a form of networking in which packet routing control can be separated from switching hardware [3]. As a result, when the SDN and Ethernet fabrics are consolidated, real network intelligence is achieved [4].

Since OpenFlow is the industrial standard interface for SDN between the control and the data layers, the following subsection defines it and its architecture.

### 2.3 Open Flow Definition

OpenFlow is the protocol used for managing the southbound interface of the generalized SDN architecture. It is the first standard interface defined to facilitate interaction between the control and data planes of the SDN architecture. OpenFlow provides software-based access to the flow tables that instruct switches and routers how to direct network traffic. Using these flow tables, administrators can quickly change network layout and traffic flow. In addition, the OpenFlow protocol provides a basic set of management tools which can be used to control features such as topology changes and packet filtering. The OpenFlow specification is controlled and defined by the non-profit open network foundation (ONF), which is led by a board of directors from seven companies that own and operate some of the largest networks in the world (Deutsche Telekom, Facebook, Google, Microsoft, Verizon, Yahoo, and NTT). Most of the networking hardware vendors such as HP, IBM, and CISCO offer switches and routers that use the OpenFlow protocol [10]. OpenFlow shares much common ground with the architectures proposed by ForCES and SoftRouter; however, the difference lies in inserting the concept of flows and leveraging the existence of flow tables in commercial switches [11].

OpenFlow-compliant switches come in two main types: OpenFlow-only and OpenFlow-hybrid. OpenFlow-only switches support only OpenFlow operations, i.e., all packets are processed by the OpenFlow pipeline. OpenFlow-hybrid switches support both OpenFlow operations and normal Ethernet switching operations, i.e., traditional L2 and L3 switching and routing. These hybrid switches support a classification mechanism outside of OpenFlow that routes traffic to either of the packet-processing pipelines [11].

### 2.3.1 OpenFlow Architecture

Basically, the OpenFlow architecture consists of numerous pieces of OpenFlow-enabled switching equipment which are managed by one or more OpenFlow controllers, as shown in Fig.5.

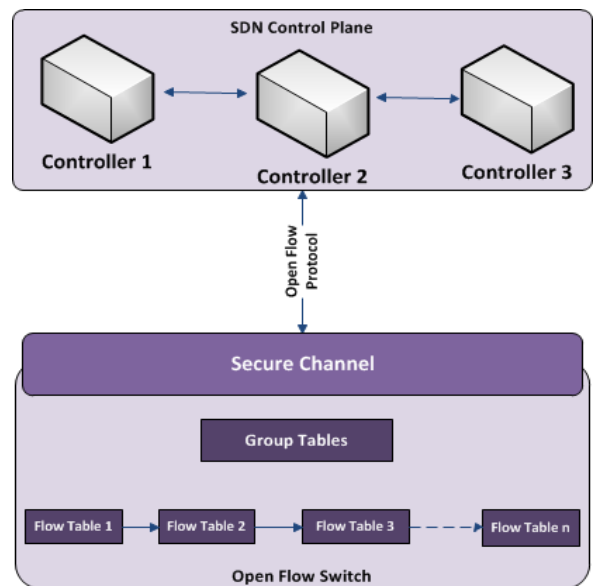

**Figure 5: Basic Architecture of OpenFlow**

*a) Defining a Flow*

Network traffic can be partitioned into flows, where a flow could be a transmission control protocol (TCP) connection, packets with the same MAC address or IP address, packets with the same virtual local area network (VLAN) tag, or packets arriving from the same switch port [9].

*b) OpenFlow Switch*

An OpenFlow switch consists of one or more flow tables and a group table. It performs packet look-ups and forwarding. The controller manages the OpenFlow-enabled switch using the OpenFlow protocol over a secure channel. Each flow table in the switch is made up of a set of flow entries in which each flow entry consists of match header fields, counters, and a set of instructions to apply to matching packets [11].

*c) OpenFlow Channel*

The OpenFlow channel is the interface that connects each OpenFlow switch to a controller. Using this interface, the controller configures and manages the switch. The OpenFlow protocol supports three message types, all of which are sent over a secure channel. These messages can be categorized as controller-to-switch, asynchronous, and symmetric, each having multiple sub-types. Controller-to-switch messages are initiated by the controller and are used to manage or derive information directly about the state of the switch. Asynchronous messages are initiated by the switch and are used to update the controller with network events and changes to the switch state. Symmetric messages are initiated by either the switch or the controller and are sent without solicitation. The OpenFlow channel is usually encrypted using transport layer security (TLS), but can also operate directly over TCP [11].

### *d) OpenFlow Controller*

The controller is responsible for maintaining all the network protocols and policies and distributing appropriate instructions to the network devices. In other words, the OpenFlow controller is responsible for determining how to handle packets without valid flow entries. It manages the switch flow table by adding and removing flow entries over the secure channel using the OpenFlow protocol. The controller essentially centralizes network intelligence. The switch must be able to establish communication with a controller at a user-configurable (but otherwise fixed) IP address using a user-specified port. The switch initiates a standard TLS or TCP connection to the controller when it knows its IP address. Traffic to and from the OpenFlow channel does not travel through the OpenFlow pipeline. Therefore, the switch must identify incoming traffic as local before checking it against the flow tables. The switch may establish communication with a single controller or with multiple controllers.

Having multiple controllers improves reliability because the switch can continue to operate in OpenFlow mode if one controller connection fails. The hand-over between controllers is entirely managed by the controllers themselves, which enables load balancing and fast recovery from failure. The controllers coordinate the management of the switch among themselves, and the goal of the multiple controller functionality is only to help synchronize controller hand-offs performed by the controllers.

The multiple controller functionality addresses only controller fail-over and load balancing. When OpenFlow operation is initiated, the switch must connect to all controllers with which it is configured and try to maintain connectivity with all of them concurrently. Many controllers may send controller-to-switch commands to the switch; the reply or error messages related to these commands must be sent only on the controller connection associated with that command. Typically, the controller runs on a network-attached server [11].

SDN controllers can be implemented in the following three structures [12]:

     i. Centralized structure

     ii. Distributed structure

     iii. Multi-layer structure.

### *2.3.2 Flow & Group Tables*

Each entry in the flow table has three fields [11]:

• A packet **header** is specific to the flow and defines it. This header is almost a ten-tuple. Its fields contain information such as VLAN ID, source and destination ports, IP address, and Ethernet source and destination.

•The **action** specifies how the packets in a flow will be processed. An action can be any one of the following:

i) Forward the packet to a given port or ports

ii) Drop the packet

iii) Forward the packet to the controller.

• **Statistics** include information such as number of packets, number of bytes, time since the last packet matched the flow, and so on for each type of flow [11]. Most of the time, counters are used to keep track of the number of packets and bytes for each flow and the elapsed time since flow initiation.

### *2.3.3 OpenFlow Protocol*

An OpenFlow switch contains multiple flow and group tables. Each flow table consists of many flow entries. These entries are specific to a particular flow and are used to perform packet look-up and forwarding. The flow entries can be manipulated as desired through OpenFlow messages exchanged between the switch and the controller on a secure channel. By maintaining a flow table, the switch can make forwarding decisions for incoming packets by a simple look-up on its flow-table entries. OpenFlow switches perform an exact match check on specific fields of the incoming packets. For every incoming packet, the switch goes through its flow table to find a matching entry. The flow tables are sequentially numbered, starting at 0. The packet-processing pipeline always starts at the first flow table. The packet is first matched against the entries of flow table 0. If the packet matches a flow entry in a flow table, the corresponding instruction set is executed. Instructions associated with each flow entry describe packet forwarding, packet modification, group table processing, and pipeline processing.

Pipeline-processing instructions enable packets to be sent to subsequent tables for further processing and enable aggregated information (metadata) to be communicated between tables. Flow entries may also forward to a port. This is usually a physical port, but may also be a virtual port.

Flow entries may also point to a group, which specifies additional processing. A group table consisting of group entries offers additional methods of forwarding (multicast, broadcast, fast reroute, link aggregation, etc.). A group entry consists of a group identifier, a group type, counters, and a list of action buckets, where each action bucket contains a set of actions to be executed and associated parameters. Groups also enable multiple flows to be forwarded to a single identifier, e.g., IP forwarding to a common next hop. Sometimes packet may not match a flow entry in any of the flow tables; this is called a "table miss". The action taken in case of a miss depends on the table configuration. By default, the packet is sent to the controller over the secure channel. Another option is to drop the packet [11].

In summary, SDN provides a new concept and architecture for managing and configuring networks using a dynamic and agile infrastructure. But the networking area is not only experiencing the emergence of SDN but also network virtualization and network function virtualization. The three solutions build an automated, scalable, virtualized and agile networking and cloud environment. Therefore, the following section introduces network virtualization, network function virtualization and their relationship with SDN.

### 3. NETWORK VIRTUALIZATION

The value of SDN in the enterprise lies specifically in its ability to provide network virtualization and automated configuration across the entire network fabric, enabling rapid deployment of new services and end systems in addition to minimizing operating cost[13][14].

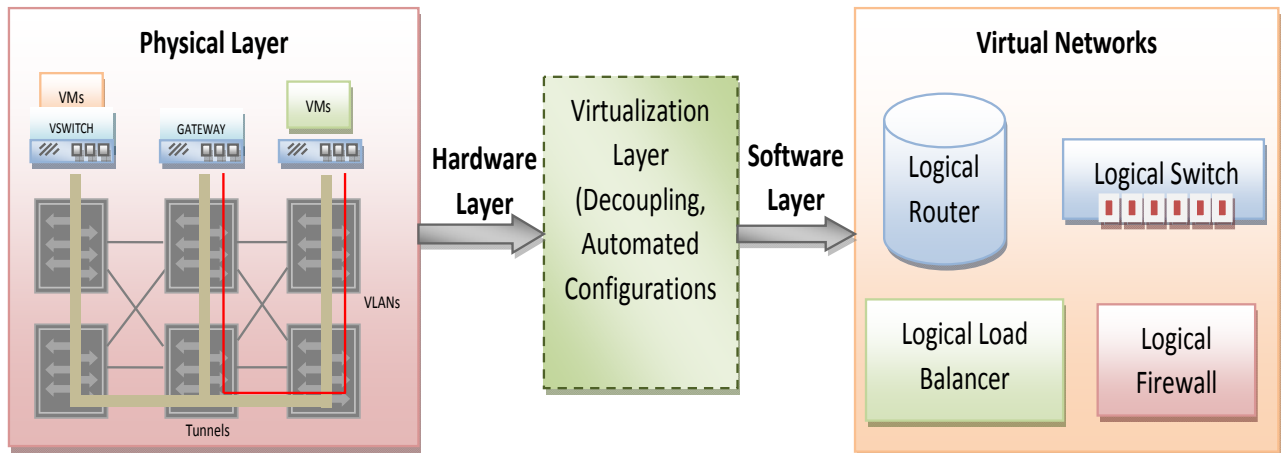

**Figure 6: The Concept of Network Virtualization**

*3.1 Definition*

Conceptually, network virtualization decouples and isolates virtual networks from the underlying network hardware, as shown in Fig.6.

The isolated networks share the same physical infrastructure. Once virtualized, the underlying physical network is used only for packet forwarding. Multiple independent virtual networks are then overlaid on the existing network hardware, offering the same features and guarantees as a physical network, but with the operating benefits and hardware independence of virtual machines [13]. To achieve virtualization at the network level, a network virtualization platform (NVP) is needed to transform the physical network components into a generalized pool of network capacity, similar to how a server hypervisor transforms physical servers into a pool of compute capacity. Decoupling virtual networks from physical hardware enables network capacity to scale without affecting virtual network operations [13, 14].

Network virtualization projects the network hardware as a business platform capable of delivering a wide range of IT services and corporate value. It delivers increased application performance by dynamically maximizing network asset utilization while reducing operating requirements [7].

With the emergence of SDN, network virtualization becomes engaged in cloud computing applications. NV provides network management for the interconnection between servers in DCs. It allows the cloud services to be dynamically allocated and extend the limits of DC into the network infrastructure [15].

Network virtualization has many aspects, including virtualized dual backbones, network service virtualization, virtual service orchestration, network I/O virtualization, and network-hosted storage virtualization. Figure 7 presents a general network-virtualization architecture consisting of firewall-defined layers [16, 17].

Starting from the bottom:

*e) Infrastructure Provider (InP):*

The infrastructure provider (InP) is responsible for maintaining the underlying physical equipment. Each organization taking on this role must offer its resources virtually to build a virtual network.

*f) Virtual Network Provider (VNP)*

The virtual network provider (VNP) is responsible for requesting virtual resources and assembling the virtual network for a virtual network operator (VNO). The virtual network provider can use a number of infrastructure providers to provide virtual network resources.

*g) Virtual Network Operator (VNO)*

VNOs must assess the network requirements for the VNP to assemble the virtual resources. VNOs are also responsible for managing and granting access to virtual networks.

*h) Service Provider*

Service providers use virtual network resources and services to tailor specialized services for end users.

*i) Virtual Network User/End User*

End users consume the resources of the virtual network through services provided by the virtual network directly or services provided by a service provider.

Three components are essential for a virtual network to function properly: virtual servers, virtual nodes, and virtual links. Virtual servers provide end users with a means to access virtual network services by implementing virtual machines. The virtual servers can also switch transparently between virtual machines to enable dynamic service changes. This feature is particularly helpful in the face of ever-changing client needs. Virtual nodes represent physical nodes such as routers and switches. A virtual node operates in both the data and control planes. The node is configured by VNOs to forward data appropriately. Virtual links provide a means of dividing and sharing physical links. The concept of virtual links ensures flexibility in network topology [16, 17].

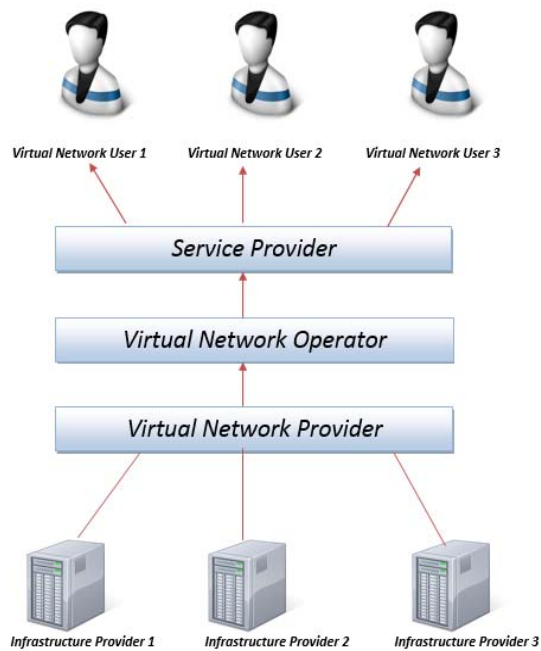

**Figure-7: General Network Virtualization Architecture**

### 3.2 Benefits of Network Virtualization

Some of the key benefits offered by network virtualization are mentioned below [18, 19]:

*1) Co-existence of Dissimilar Networks*

Network virtualization makes it possible to create multiple virtual networks on the same physical hardware. However, these virtual networks can be isolated from other existing virtual networks. This isolation can be used as a tool in the deployment of networks using different or even incompatible routing protocols.

*2) Encouraging Network Innovation*

Like SDN, network virtualization can be used to encourage innovation in the networking domain. The isolation that can exist between two virtual networks can be used to create separate domains for production traffic and test traffic. This isolation guarantees that a malfunction experiment will not affect production traffic.

*3) Provisioning of Independent and Diverse Networks*

NV deploys packet handling, quality of service (QoS) and security policies to configure network operations and behaviors. This configuration allows the categorization of different networks based on their services, users and applications.

*4) Deployment of agile network capabilities*

The inclusion of agile facilities into the current network improves the data transport efficiency and provides robust network. With the agile manner, NV allows the integration between legacy and advanced networks. Also, it enables migration from legacy systems into advanced ones in an agile manner.

*5) Resource Optimization*

The dynamic mapping of multiple virtual network nodes to the physical substrate ensures that the network hardware is utilized up to capacity. This approach cuts down on hardware costs and delivers additional profit to the infrastructure provider.

*6) Deployment of Distinct Network Services*

Network services such as wireless local area networks (WLANs) and Intranet require specific network architectures. In addition, a multi-national corporation might need to offer distinct services to its employees. This can add complexity to the existing overlay network. Network virtualization can help alleviate these problems by deploying such services in separate virtual networks.

### 3.3 Network Function Virtualization

As for the ambiguity between the concepts of network function virtualization (NFV) and SDN, it is necessary to take advantage of the definitions and benefits of both technologies.

#### 3.3.1 Definition of NFV

Expansion of the deployment of various applications and network services induced service providers to come up with the concept of NFV. Therefore, they established a European telecommunication standards institute (ETSI) Industry Specification Group for NFV. The group defined the real concept of NFV together with its requirements and architecture.

NFV decouples network functions, e.g., firewalls, domain name service (DNS), and caching, from dedicated hardware appliances and entrusts them to a software-based application running on a standardized IT infrastructure, high-volume servers, switches, and storage devices. The interesting feature of NFV is its availability for both wired and wireless network platforms. NFV reduces capital expenditures (CAPEX) and operating expenditures (OPEX) by minimizing the purchase of dedicated hardware appliances, as well as their power and cooling requirements. Virtualization of network functions enables fast scale-up or scale-down of various network services and provides agile delivery of these services using a software application running on commercial off-the-shelf (COTS) servers.

#### 3.3.2 NFV and NV

At the flow level, NV partitions the network logically and creates logical segments in it. It can be viewed as a tunnel connecting any two domains in a network. Therefore, it eliminates the physical wiring for each domain connection and replaces it with a virtualized infrastructure.

While NV can be viewed as a tunnel, NFV deploys services on it. NFV virtualizes the functions from layer 4 (L4) till layer 7 (L7) such as load balancing and firewalls. If an administrator can enable modifications at the top of the infrastructure layer, NFV can provide changes for L4-L7 functions virtually

Both NV and NFV may run on high performance x86 platforms. NV tunnel facilitates VM migration independently of the underlying network. NFV enables the functions on NV, provides an abstraction and virtual services on it. As NV eliminates the need for network reconfiguration, NFV saves time on manual training and provisioning.

|  | **SDN** | **NFV** |
|---|---|---|
| **Motivation** | • Decoupling of control and data planes<br>• Providing centralized controller and network programmability | Abstraction of network functions from dedicated hardware appliances to COTS servers |
| **Network Location** | Data centers | Service provider networks |
| **Network Devices** | Servers and switches | Servers and switches |
| **Protocols** | OpenFlow | N/A |
| **Applications** | Cloud orchestration and networking | Firewalls, gateways, content delivery network |
| **Standardization Committee** | Open Networking Forum (ONF) | ETSI NFV group |

Table I: Comparison between SDN and NFV.

*3.3.3 NFV and SDN*
SDN and NFV are complementary technologies; they do not depend on each other. However, both concepts can be merged to mitigate potentially the challenges of legacy networks. The functions of the SDN controller can be deployed as virtual functions, meaning that the OpenFlow switches will be controlled using NFV software. The multi-tenancy requirements of the cloud pushed the NFV to support use of a software overlay network. This software network is created by SDN. It consists of a set of tunnels and virtual switches that prohibits sudden interactions between different virtual network functions. These functions will be managed using the SDN model.

Merging NFV and SDN enables replacement of expensive and dedicated hardware equipment by software and generic hardware. The control plane is transferred from dedicated platforms to optimized locations in DCs. Abstraction of this plane eliminates the need to upgrade network appliances simultaneously and thus accelerates the evolution and deployment of network services and applications. Table I provides a comparison between SDN and NFV concepts.

In summary, while NV and NFV creates virtual tunnels and functions to the underlying physical network respectively, SDN modifies the physical network. Also, NV and NFV can reside on the servers of the existing network, SDN requires the construction of new network where the control and data layers are decoupled.

## 4. SDN APPLICATIONS

SDN is a promising approach that can overcome the challenges facing cloud computing services, specifically NaaS and DCNs. Therefore, the following section highlights the importance of SDN in these fields and describes its various applications in DCNs and NaaS.

*4.1 Data-Center Networks*
*4.1.1 Motivation*
The scale and complexity of data-center networks (DCNs) are approaching the limit of traditional networking equipment and IT operations [20]. Currently, the infrastructure of data-center networks is undergoing tremendous and rapid changes.
The Enterprise Strategy Group (ESG) has defined the reasons that have provoked these changes and summarizes them as follows:

- *Aggressive Alliances in Data Centers*

ESG's research statistics show that 63% of the enterprises polled are planning the fusion of their data centers [20]. A large expansion may occur in these data centers as they harbour extra applications, network traffic, and devices. Therefore, many associations might consolidate their data centers into multi-tenant facilities.

- *Progressive use of virtualization technology*

Large enterprises such as Citrix, Microsoft, and VMware are deploying server virtualization technologies. In addition, other organizations are now willing to introduce new initiatives to their infrastructure that use virtualization technology concepts. Consequently, compact integration among physical infrastructure, virtual servers, and networks is required.

- *Wide Deployment of Web Applications*

Web-based applications are widely used in many organizations [20]. Moreover, these applications use server-to-server communication because they are based on x86 server tiers and horizontal scaling. Therefore, data centers need to brace themselves for an increase in internal traffic due to massive deployment of these Web applications.

| Changes in DCN Infrastructure |||
| --- | --- | --- |
| **Proposed Solution** | **Objective** | **Functionality** |
| SDN-based Vello Systems [22] | • Override the traditional Layer 2 domains and Layer 3 routing challenges<br>• Facilitate live VM migration within and across DCNs | • Enable migration of performance and QoS and security policies with VM migration<br>• Provide a unified view and control of the global cloud for WAN resource optimization<br>• Provide network automation and virtualization of LAN and WAN connectivity and resource allocation |
| Switching with in-packet Bloom filters (SiBF) [23] | • Transform the DCN into a software problem<br>• Leave the responsibility for device implementation to hardware vendors | • Override the single point of failure problem by using a distributed controller system<br>• Provide load-balancing services<br>• Guarantee better scalability and fault tolerance performance in DCN by using rack managers<br>• No evaluation of the proposed routing approach<br>• Lack of traffic engineering studies of different flow sizes |

Table II: SDN in DCN Infrastructure.

Because dynamic scaling in data-center networks is based on static network devices (Ethernet and IP packet connections), IT teams encounter a discontinuity gap during the implementation of scalable data-center networks. However, it appears that the flood waters are about to overrun tactical network sandbags [20]. The ESG describes the main network challenges as follows:

*a) Network Segmentation and Security*

Nowadays, DCN segmentation is based on a mix of VLANs, IP subnets, device-based access-control lists (ACLs), and firewall rules that have been maintained for years. However, these hard-wired segmentation and security controls are not compatible with data centers that are populated by VM workloads and cloud-computing platforms.

*b) Traffic Engineering*

Any traffic congestion or hardware failure will affect the performance and latency of all other devices because network traffic follows fixed paths and multiple hops. In addition, the deployment of VMs and virtual servers in recent DCNs adds a supplementary burden to network performance [21].

*c) Network Provisioning and Configuration*

Although virtual servers are provisioned by cloud orchestration tools, the policies of the data-center equipment and control paths must be set up on a device-to-device or flow-to-flow basis, and heterogeneous networks must be managed by multiple management systems. Even though network management software can help at this stage, network configuration changes remain "a tedious link-level slog" [20]. Further information concerning DCN challenges can be found in [20, 21]. Ultimately, DCN discontinuity will be a threat to business operations because it may induce degradations in service level, delays in business initiatives, and increase in IT operating costs [20]. Although networking vendors have launched some innovations such as network fabric and convergence architectures to fix the fractures in the DCN infrastructure, these solutions do not address the problems in heterogeneous networks. Nevertheless, the software-defined network paradigm is a promising solution to solve these challenges in DCN setups.

*4.1.2 SDN Deployment in DCNs*

In SDN OpenFlow based-networks, the virtual network segments are centrally configured, and network security is simplified by directing flows to security policy services. Moreover, the central controller transforms the core and aggregation devices into a "high-speed transport backplane" [20]. The controller can provision a new device that is added to the network and allow it to receive the configuration policy when it appears online. Finally, SDN improves DCN infrastructure, its power consumption, and its various metrics. Due to these improvements and modifications, different SDN applications in DCNs have been proposed.

*a. Changes in DCN Infrastructure*

Automation and virtualization of data-center LANs and WANs has resulted in a flexible and dynamic infrastructure that can accommodate operating-cost challenges. As a result, Vello Systems [22] has proposed an open and scalable virtualization solution that connects the storage and computation resources of the data center to private and public cloud platforms. To facilitate the migration of VMs from their Layer 2 network, Layer 2 was extended across multiple DCs using Layer 3 routing. However, Layer 3 routing introduces challenges in intra-data center connectivity and cannot meet the requirements for VM migration across DCs. Therefore, the proposed solution is based on a cloud-switching system that enables cloud providers and enterprises to overcome the traditional Layer 2 domains, the direct server-to-server connection, and virtual server migration.

Because the switching system supports integration of end-to-end network attributes, its operating system can provide a framework for SDN. Thus, OpenFlow-based allow the cloud to migrate performance, QoS, and security policies concurrently with VM migration. Finally, the SDN-based Vello systems permit a unified view and control of the global cloud for WAN resource optimization.

| Green DCN |||
|---|---|---|
| **Proposed Solution** | **Objective** | **Functionality** |
| OpenFlow platform for energy-aware data center [24] | Provide guidelines for studying energy consumption in DCN elements | • Estimate the minimum power for a given network topology<br>• Satisfy the traffic conditions and QoS requirements<br>• Provide a power module in the controller that determines the power state of network elements<br>• No evaluation of the proposed approach on different network topologies |
| OpenFlow switch controller (OSC) [26] | Decrease the influence of carbon emissions in the DCs | • Reduce configuration time of network elements<br>• Enable flexible power management operations based on the programmable controller |

Table III: SDN in a Green DCN.

In [23], an OpenFlow-based test-bed implementation, switching with in-packet Bloom filters (SiBF), has been proposed as data-center architecture. The suggested architecture was inspired by the onset of SDN, which transforms the DCN into a software problem while leaving the hardware vendors responsible for device implementation. SiBF introduces an army of rack managers that act as distributed controllers, contain all the flow-setup configurations, and require only topology information. Intrinsically, SiBF uses IP addresses for VM identification and provides load-balanced services based on encoding strategies. The architecture is implemented on a multi-rooted tree (CORE, AGGR, and ToR) because this is a common DCN topology.

However, other topologies can be considered in a SiBF datacenter architecture. The OpenFlow controller, e.g., the rack manager, installs the flow mapping into the ToR switches and consists of directory services, topology services, and topology discovery. With its modules, the controller can be implemented as an application in the NOX controller. Flow requests are handled by neighbouring rack managers in case of any failure in the master controller. However, when an OpenFlow switch fails, its traffic is interrupted until the SiBF installs new mappings (new flow routes) in the ToR switches. The proposed data-center architecture, based on distributed OpenFlow controllers, guarantees better scalability and faulttolerant performance in the DCN. Table II summarizes various approaches for implementing SDN in a DCN infrastructure.

*b. The Green DCN*

Implementing an energy-efficient data-center network is an important step towards a "green" cloud. An energy-aware data-center architecture based on an OpenFlow platform has been proposed in [24]. Because designing an energy-efficient data center requires an experimental environment, the authors in [24] analyzed the proposed architecture based on the Reducing Energy Consumption in Data-Center Networks and Traffic Engineering (ECODANE) project. The platform provides guidelines for measuring and analyzing energy consumption in DCN elements (ports, links, and switches) based on realistic measurements from NetFPGA-based OpenFlow switches. The NetFPGA energy model was extracted from several energy measurements using the Xilinx power-estimation tool. The Xpower tool uses the Verilog source code of the OpenFlow switch as its input and estimates the power measurements.

The power-estimation model was tested using the Mininet [25] emulator, a simple testbed for developing OpenFlow applications. The Elastic Tree topology was used to test the proposed data-center architecture. The OpenFlow switches are controlled by the NOX controller, which consists of an optimizer, a power controller, and routing modules. The optimizer finds the minimum power of the network subset that satisfies the traffic conditions and QoS requirements. The minimum power estimate is deduced from the number of links or switches that are turned off or put in sleep mode. The power-control module determines the power state of the network elements based on the OpenFlow messages and Mininet APIs and notifies switches to enter the appropriate power-saving mode. The last module is used to find the optimal routing path in the DCN. This study is a first stage in building a green data center based on the new SDN paradigm. Table III summarizes various SDN approaches in a green DCN.

Based on the results and the proposed data-center architecture described in [24], an extension to OpenFlow switches for saving energy consumption in data centers has been proposed in [26] and can be used later on as a reference. The authors presented a solution to decrease the environmental influence of massive carbon emissions in data centers. This solution consists of controlling power consumption in data-center switches based on an extension to OpenFlow switches. This extension adds new messages to the OpenFlow protocol which enable the controller to control the switch over different power-saving modes. More detailed information about the new power-control messages and the design of the OpenFlow Switch Controller (OSC) can be found in [26]. OpenFlow can reduce configuration time and enable flexible programmable controller-based management operations and is therefore recommended for use in cloud data centers.

| DCN Metrics |||
|---|---|---|
| **Proposed Solution** | **Objective** | **Functionality** |
| OpenFlow platform for scalable and agile data center [25] | • Improve performance, scalability, and agility in a cloud data center | • Improve bandwidth performance and the number of flow modifications per second in the kernel switches<br>• Reduce cost of operations and switch configuration time |
| Loss-free multipathing congestion control DCN [28] | • Provide lossless delivery and better throughput for DCN using OpenFlow switches and a central controller | • Reduce path-load update overhead of the network<br>• Handle any network status and traffic burst states<br>• Use the path load as the only parameter to evaluate traffic in the DCN |
| SDN-based DCN solution [30] | • Meet the requirements of different applications in a cloud DCN | • Remove the limitation on the number of VLANs<br>• Respond to on-demand network updates<br>• Introduce longer flow-setup time compared to legacy networks |
| OpenFlow re-routing control mechanism in DCN [32] | • Evaluate DCN performance and manage its flows | • Use the least loaded route and alternative paths for flow congestion<br>• Provide storage of switch statistics, tracking all the detected hosts in the network and various routing and re-routing functions<br>• Provide better load distribution, throughput, and link utilization compared to other routing mechanisms<br>• Combat severe packet loss and high packet sojourn time in case of low processing time for private networks |
| Scissor [33] | • Modify packet headers to minimize DC traffic and network power consumption | • Replace redundant header information with a short identifier, the Flow ID<br>• Combine packets of the same flow in the same ID<br>• Improve latency and introduce slight improvements in power gains<br>• Absence of scissor operations within the rack that is responsible for 75% of DCN traffic |

Table IV: Improvements in DCN Metrics with SDN.

*c. Improving DCN Metrics*

[27] describes an experimental study for improving the performance, scalability, and agility of a cloud data center using the OpenFlow protocol. The authors built a prototype cloud data center in which the route traffic was controlled by an OpenFlow controller; different metrics were tested on the prototype. The proposed algorithms and the prototype design are discussed in detail in [27]. Testing of performance, throughput, and bandwidth for various network sizes and topologies was done using the Mininet emulator with its numerous tools. The results show that bandwidth performance and the number of flow modifications per second were better with the Kernel switches, a test image of OpenFlow, than with user-space switches. However, replacement of data-center switches with OpenFlow switches is not recommended until standardization of the software platform has been achieved.

Furthermore, SDN has often been mentioned as an approach to implementing and improving the metrics of data-center networks. In [28], a loss-free multipathing (MP) control congestion (CC) approach for a DCN was proposed. The authors introduced integration between MP and CC to provide lossless delivery and better throughput for the DCN.

The integration mechanism was based on a dynamic load-balancing multipathing approach [29]. The proposed mechanism uses OpenFlow switches and a central controller to reduce network overhead (path load updates) and enable the switches to deal with any network situation even during traffic bursts [28]. OpenFlow is enabled only in the access switches. The controller collects information about the network from their routing tables. The controller updates the switches with any change in the "path load" on the associated routes with a short delay. Although the MP-CC integration mechanism shows lossless delivery due to fast reaction of the switches to network changes, the proposed algorithm considers path load as the only parameter to handle DCN traffic.

Recent applications have imposed many requirements on cloud service providers, and therefore, cloud data-center networks have to be multi-tenant, low-cost, flexible, and reconfigurable on demand. On the other hand, current DCN strategies cannot meet all these requirements, and therefore [30] proposed an SDN-based network solution.

The proposed prototype consists of a central controller that manages multiple OpenFlow switch instances and packet filtering. The controller stores the database of the management information of the L2 virtual network, called the slice. The proposed prototype removes the limitations on the number of VLANs and responds to on-demand network updates based on APIs that simplify these configuration updates. However, the flow-setup process in the switch introduces a longer flow-setup time than in legacy networks [30].

| Virtualized DCN |||
|---|---|---|
| **Proposed Solution** | **Objective** | **Functionality** |
| Inter-DCN connectivity based on OpenFlow [34] | Mitigate the interconnection challenges in a cloud DCN | • Insert new OpenFlow rules to implement inter-DCN connectivity in the cloud<br>• Support live VM migration between different DCNs<br>• Minimize connectivity interruption of VMs during the migration process |
| CrossRoads [35] | Facilitate live and offline VM migration across data centers | • Support East-West traffic for migration within DCs<br>• Support North-South traffic for VM migration to external clients<br>• Provide negligible overhead with respect to that of legacy networks |

Table V: Virtualized DCN using SDN.

Another study proposed an approach to evaluate DCN performance by implementing an OpenFlow re-routing control mechanism to manage DCN flows [31]. Performance is represented by load distribution, throughput, and link utilization metrics. The proposed re-routing scheme initially uses the least loaded route; in case of congestion, large flows are re-routed onto alternative paths, while small flows pursue their track. The re-routing framework consists of an NOX controller, a monitor to store switch statistics, a host tracker that tracks the entire set of detected hosts in the network, and finally a routing engine which is responsible for routing and re-routing functions.

A comparison between the single-path, equal-cost multi-path, and OpenFlow re-routing mechanisms showed that the proposed framework has better load distribution, throughput, and link utilization. Table IV summarizes the improvements in DCN metrics using the SDN approach.

In spite of the benefits provided by introducing SDN into DCNs, [32] concluded that building an OpenFlow system requires observation of the relative load on the OpenFlow controller. The authors studied the performance of this controller in educational and private networks and concluded that a processing time of 240 μs is sufficient for an educational network, but that private networks require a more powerful controller with better processing time or distributed controllers; otherwise, severe packet loss and high packet sojourn times may occur [32].

The OMNet++ simulation environment was used to evaluate system performance by measuring relative packet loss and mean packet sojourn time.

Packet headers are responsible for 30%–40% of DC traffic [33] and network power consumption. Therefore, the authors of [33] proposed a new framework, the "Scissor", which replaces redundant header information with a short identity, the "Flow ID".

The Flow ID identifies all the packets belonging to the same flow. Trimming of header information is done by micro-architectural hardware that consists of multiplexers to select the fields that will be retained by the Scissor, a buffer to hold the complete header temporarily, ternary content-addressable memory (TCAM), and a controller that generates the Flow IDs. Experimental simulations were carried out to test the performance of the proposed framework. Results showed that Scissor introduced substantial latency improvements, as high as 30%.The evaluated power gains were only 20% in the best-case scenario because no scissor operations were performed within the rack that is responsible for 75% of the DCN traffic [33].

*d. Virtualization in DCNs*

The SDN approach mitigates the interconnection challenges of cloud DCNs [32]. The characteristics of heterogeneous DCN architectures (VL2, Portland, and Elastic Tree) are represented by OpenFlow rules. These rules are passed to all DCN elements to implement "inter-DCN connectivity" [34]. These rules support VM migration between different DCN schemes without connectivity interruption based on re-routing mechanisms.

Live VM migration in DCN is crucial in the case of disaster recovery, providing fault tolerance, high availability, dynamic workload balance, and server consolidation [35]. This reference proposed a network fabric based on OpenFlow, "CrossRoads", that enables both live and offline VM migration across data centers. CrossRoads supports East-West traffic for VM migration within data centers and north-south traffic for VM migration to external clients. The framework consists of a centralized controller in each data center, thus extending the controller placement problem. Table V presents a couple of implemented SDN approaches to virtualized DCNs.

Experimental results showed that the proposed network fabric has negligible overhead with respect to the default network and outperforms the default network by 30% [35].In summary; SDN is a promising solution that alleviates most of the challenges faced by cloud DCNs. However, recent research studies have been based on small topologies or emulators. Therefore, coupling SDN to a DCN and a cloud resource environment and testing the performance of the scheme on a real large network is needed to achieve better understanding of the performance of SDN-based DCN setups.

## 4.2 Network as a Service

### 4.2.1 Service Oriented Architecture

The service oriented architecture (SOA) is the concept of building a software system based on multiple integrated logical units. These units known as services allow better construction and management to solve large problems in different domains. The basic components of SOA are elucidated in Fig.8. The architecture depends on the Services used by the Service User entity. The Service Provider hands over these services and the Service Registry coordinates the services' information and publishes them for the Service User [36].

SOA satisfies the requirements of different applications by balancing the computational resources. It virtualizes and integrates these resources in form of services entities. Therefore, the basic aspect of SOA is the "coupling" between different systems. Every system has information about the behavior and implementation of its partners. The information gathering procedure facilitates the coupling feature in SOA.

SOA eliminates the tight coupling and lack of interoperability between diverse middleware in a network. It has been endorsed by Cloud Computing (CC) services; Infrastructure as a Service (IaaS), Platform as a Service (PaaS) and Software as a Service (SaaS). Fig.9 elucidates SOA in CC environment. CC implements SOA in its different fields to exploit the resources' virtualization feature. This in turn allows SOA to introduce Network as a Service (NaaS) into CC [7].

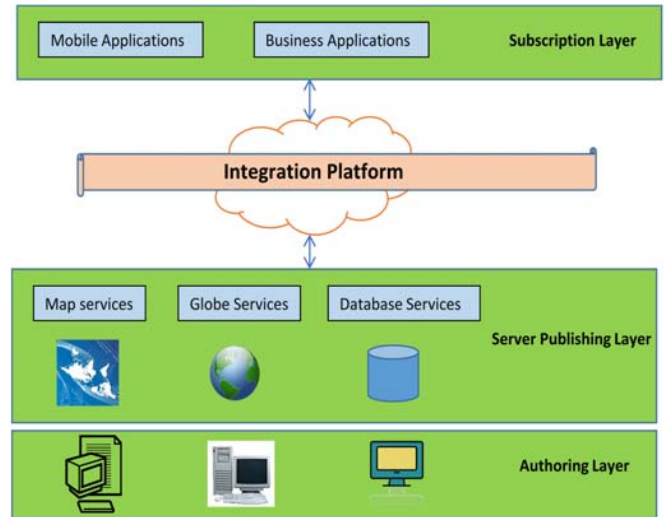

**Figure-9: SOA in Cloud Computing Environment**

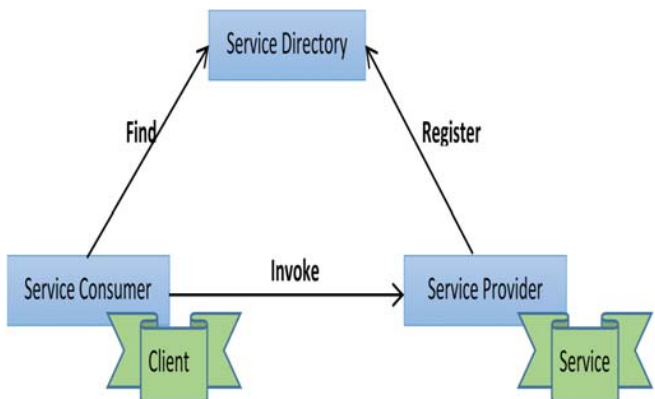

**Figure-8: Components of SOA**

### 4.2.2 Motivation

Cloud computing offers on-demand provisioning of computational resources for tenants using a pay-as-you-go model and outsources hardware maintenance and purchases [37]. However, these tenants have limited visibility and control over network resources and must resort to overlay networks to complete their tasks. The separation of computation from end-to-end routing in traditional networks in the cloud-computing environment could affect data-plane performance and control-plane flexibility.

These drawbacks can be addressed by NaaS. It provides secure and direct access for tenants to cloud resources and offerings and enables efficient use of the network infrastructure in the data center [37]. NaaS is a new Internet-based model that enables a communication service provider to provide network resources on demand to the user according to a service-level agreement (SLA). NaaS can also be seen from the service point of view as an abstraction between network functions and protocols [38].

The top abstraction layers deal with NaaS as a service that uses the network and customizes its capacity. Customization in the lower layers is replaced by resource management policies. [13] defines Naas as *Telco as a Service* (TaaS), which offers a "common network and IT substrate that can be virtualized and combined as a slice".

Naas is also defined as a Web 2.0 model that provides software as a service utility by exposing network capabilities (billing, charging, location, etc.) as APIs to third-party application service providers [39]. In NaaS, the owners of the underlying network infrastructure offer virtual network services to a third party. There is a clear demarcation between the roles of infrastructure providers (InPs) and virtual network operators (VNOs).

The InP is responsible for the operating processes in the underlying network infrastructure, and the VNO is responsible for the operating processes in the virtual networks that run on top of the physical infrastructure.

The NaaS scenario offers many business incentives, such as higher revenues for InPs and lower capital and operating expenditures for VNOs, because it enables a number of virtual networks to run on the same underlying network infrastructure. Detailed information on the interaction between InPs and VNOs is available in [13].

In summary, NaaS provides the following benefits to operators [40]:
- Faster time to transition NaaS to market.
- Self-service provisioning.
- Flexibility in upgrading NaaS resources without long-term constraints.

- Payment only for resources used.
- Repair and maintenance are part of the service.
- Greater control in adding, changing, and deleting services.

*4.2.3  NaaS and SDN Integration*

NaaS is one of the promising opportunities for SDN. NaaS providers can use SDN orchestration systems to obtain a powerful user interface for controlling and viewing network layers. A variety of research studies have proposed NaaS platforms in an SDN environment.

    *a.  Cloud-NaaS Model*

[38] introduced a cloud-based network architecture which evaluates the provision, delivery, and consumption of NaaS. The proposed cloud-based network consists of four layers: the network resource pool (NRP), the network operation interface (NOI), the network run-time environment (NRE), and the network protocol service (NPS).

The NRP consists of network resources: the bandwidth, queues, and addresses for packet forwarding. The NOI is a standardized API for managing and configuring the NRP. The NRE is the environment that performs billing, resource allocation, interconnection, and reliability assurance for protocol service instances through service migration in cases of network failures and high load [38]. Finally, the NPS is responsible for describing, managing, and composing the new implemented network protocols.

The proposed architecture is implemented using the OpenFlow protocol. The implementation consists of two layers: the controller control plane and the network data plane. The first layer is responsible for NRE and NPS functions. It consists of a master controller that distributes the data stream to the slave servers and slave controllers that perform switching, routing, and firewall functions. The data-plane layer contains the OpenFlow switches that perform packet forwarding services based on controller instructions. The authors in [38] presented NaaS in a cloud-based network, but performance and evaluation studies of the suggested implementation were not carried out.

The limitations on tenants in controlling and configuring networks in current cloud environments provided a motivation for the authors of [41] to implement a CloudNaaS model. The proposed networking framework enables the tenant to access functions for virtual network isolation, addressing, and deployment of middlebox appliances [42] for caching and application acceleration.

The CloudNaaS consists of the cloud controller and the network controller. The cloud controller manages virtual resources and physical hosts and supports the APIs which set network policies. It also specifies user requirements and transforms them into a communication matrix that resides on the OpenNebula framework. These matrices are compiled into network-level rules by the network controller (NOX controller).

The network controller installs these rules in virtual switches, monitors and manages the configuration of network devices, and decides on the placement of VMs in the cloud. The authors proposed optimization techniques to mitigate the hardware limitations mentioned in [41]. These techniques were implemented in the network controller and were designed to optimize traffic during VM placement and forwarding-entry aggregation using the same output ports. The implemented CloudNaaS exhibited good performance with an increasing number of provisioning requests and used cloud resources in an effective manner.

    *b.  Network Management in NaaS*

[43] presents a scalable graph-query design, NetGraph, which supports network-management operations in NaaS modules. NetGraph is implemented on a software module on a SDN platform. The network controller consists of multiple service modules that collect information about the physical and virtual network infrastructure.

The NetGraph module resides in the centralized controller, collects information about network topology to calculate the graph of the existing topology, and supports the service modules (NaaS modules) in their query mission. Details on the implementation design and the algorithms used (Dijkstra, TEDI, and APSP) for finding the shortest paths in a weighted graph are addressed in [43]. The authors showed that the proposed algorithms have practical compute time and are suitable for centralized architectures.

NaaS can be seen as the ultimate connection between SDN and cloud computing. NaaS is a supplementary scheme for SDN; while SDN is responsible for packet forwarding and network administration, NaaS provides application-specific packet processing for cloud tenants [42]. With NaaS schemes, the operators can control the bandwidth, routing, and QoS requirements of their data. Eventually, with SDN, operators can leverage current NaaS initiatives and build their own SDN infrastructure [44]. However, integration with existing hardware and software systems and providing diverse and efficient APIs are crucial requirements for adopting the SDN and NaaS concepts [40].

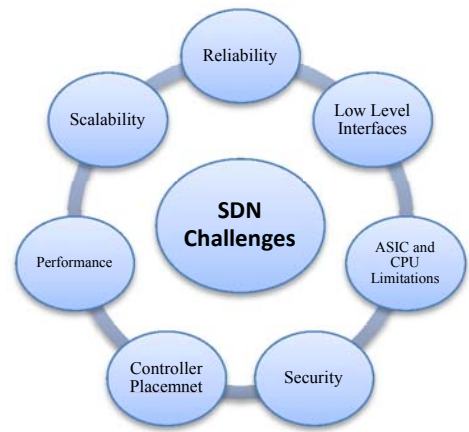

**Figure 10: SDN challenges.**

Although the SDN concept is attracting the attention of IT organizations and networking enterprises and has various applications in DCNs and NaaS, the overall adoption of SDN has encountered various obstacles, such as reliability, scalability, latency, and security challenges. Section V describes these challenges and presents some of the recent solutions proposed in the literature. Overcoming these challenges might assist IT organizations and network enterprises to improve the various opportunities and services offered by SDN.

## 5. SDN CHALLENGES AND EXISTING SOLUTIONS

Although SDN is a promising solution for IT and cloud providers and enterprises, it faces certain challenges that could hinder its performance and implementation in cloud and wireless networks [45]. Below, a list of SDN challenges and some of their existing solutions are discussed and illustrated in Fig.10.

### 5.1 Reliability

The SDN controller must intelligently configure and validate network topologies to prevent manual errors and increase network availability [46]. However, this intelligence can be inhibited because of the brain-split problem that makes the controller liable to a single point of failure [47, 48].

In legacy networks, when one or more network devices fail, network traffic is routed through alternative or nearby nodes or devices to maintain flow continuity. However, in centralized controller architecture (SDN) and in the absence of a stand-by controller, only one central controller is in charge of the whole network. If this controller fails, the whole network may collapse. To address this challenge, IT organizations should concentrate on exploiting the main controller functions that can increase network reliability [46]. In case of path/link failure, the SDN controller should have the ability to support multiple-path solutions or fast traffic rerouting into active links.

If the controller supports technologies such as Virtual Router Redundancy Protocol (VRRP) and Multi-Chassis Link Aggregation Group (MC-LAG), these might contribute to increasing network availability. In case of controller failure, it is important that the controller can enable clustering of two or more SDN controllers in an active stand-by mode; however, memory synchronization between active and stand-by controllers must be maintained [46].

The authors in [23] showed that centralized controller architecture will interrupt network traffic and flow requests in case of controller failure. Specifically, they proposed a distributed architecture, SiBF, which consists of an army of rack managers (RMs), one per rack, acting as controllers. Consequently, when the master controller fails, flow requests are handled by another stand-by controller (RM) until the master controller comes back up. In case of switch failure, SiBF installs new mappings (new back-up flow entries) in the ToR switches for each active entry. The packets in the ToR will be routed to their destinations on the alternative paths indicated by the back-up entries. Another suggested solution that can counteract reliability limitations in a centralized architecture is described in [28]. The integration between free multipathing and control congestion is based on a dynamic load-balancing multipathing approach which runs a distributed algorithm in case of controller failure. The algorithm updates the switches with any changes in "path load" on the associated routes in cases of traffic congestion and load imbalance.

### 5.2 Scalability

Decoupling between the data and control planes distinguishes SDN from a traditional network. In SDN, both planes can "evolve independently" as long as APIs connect them [49], and this centralized view of the network accelerates changes in the control plane. However, decoupling has its own drawbacks. Besides the complexity of defining standard APIs between both planes, scalability limitations may arise. Voellmy et al. [50] concluded that "when the network scales up in the number of switches and the number of end hosts, the SDN controller can become a key bottleneck".

As the bandwidth and the number of switches and flows increase, more requests will be queued to the controller, which may not be able to handle them all. Studies on a SDN controller (NOX) have shown that it can handle 30K requests/s [51]. This may be sufficient for enterprises and campus networks, but it is a bottleneck for data-center networks with high flow rates. In addition, [51] estimates that a large data center consisting of 2 million virtual machines may generate 20 million flows per second. However, current controllers can support approximately $10^5$ flows per second in the optimal case [52, 23]. In addition to controller overload, the flow-setup process may impose limitations on network scalability.

Flow setup consists of four steps:
1- A packet arrives at a switch and does not match any flow entry.
2- The switch sends a request to the controller to get instructions on how to forward the packet.
3- The controller sends a new flow entry with new forwarding rules back to the switch.
4- The switch updates its entries in the flow table.

The performance of the setup process depends on switch resources (CPU, memory, etc…) and controller (software) performance. The update time of the switch's forwarding information base (FIB) introduces delay in setting up any new flow. Early benchmarks on SDN controllers and switches showed that the controller could respond to a flow-setup request within one millisecond, while hardware switches could "support a few thousand installations per second with a sub-10 ms latency at best" [49].

Flow-setup delays may pose a challenge to network scalability. Furthermore, network broadcast overhead and the proliferation of flow-table entries limit SDN scalability [46]. The SDN platform may cause limited visibility of network traffic, making troubleshooting nearly impossible. Prior to SDN, a network team could quickly spot, for example, that a backup was slowing the network down. The solution would then be to reschedule the backup to a less busy time.

Unfortunately, with SDN, only a tunnel source and a tunnel endpoint with User Datagram Protocol (UDP) traffic are visible, but crucially, one cannot see who is using the tunnel. There is no way to determine whether the problem is the replication process, the email system, or something else. The true top talker is shielded from view by the UDP tunnels, which means that when traffic slows and users complain, pinpointing the problem area in the network is a challenge. With this loss of visibility, troubleshooting is hindered, scalability limitations emerge, and delays in resolution could become detrimental to the business [48, 53]. In order to minimize the proliferation of flow entries, the controller should use header rewrites in the network core. The flow entries will be at the ingress and egress switches.

Improved network scalability can also be ensured by enabling VM and virtual storage migration between sites, as in IaaS software middleware based on OpenFlow and "CrossRoads", a network fabric based on OpenFlow, which was discussed in previous sections [34, 35]. Another solution to scalability concerns is proposed in "DIFANE" [54]. This is a distributed flow-management architecture that can scale up to meet the requirements (large numbers of hosts, flows, and rules) of large networks.

A viable solution to scalability challenges is proposed in the "CORONET" fault-tolerant SDN architecture, which is scalable to large networks because of the VLAN mechanism installed in local switches [55]. CORONET has fast recovery from switch or link failures, supports scalable networks, uses alternative multipath routing techniques, works with any network topology, and uses a centralized controller to forward packets. It consists of modules responsible for topology discovery, route planning, traffic assignment, and shortest-route path calculation (the Dijkstra algorithm). The main feature of CORONET is the use of VLANs, which can simplify packet forwarding, minimize the number of flow rules, and support scalability properties.

In another solution, "DevoFlow" [56,57], micro-flows are managed in the data plane and more massive flows in the controller, meaning that controller load will decrease and network scalability will be maximized. This approach minimizes the cost of controller visibility associated with every flow setup and reduces the effect of flow-scheduling overhead, thus enhancing network performance and scalability.

Finally, [50] describes a scalable SDN control framework, McNettle, which is executed on shared-memory multicore servers and based on Nettle [58]. Experiments showed that McNettle could serve 5000 switches with a single controller with 46 cores and could handle 14 million flows per second with latency below 200 μs for light loads and 10 ms for loads consisting of up to 5000 switches [50].

*5.3 Performance under Latency Constraints*

SDN is a flow-based technique, and therefore its performance is measured based on two metrics: flow-setup time, and the number of flows per second that the controller can handle [46]. There are two ways to setup a flow: proactive and reactive. In proactive mode, flow setup takes place before packet arrival at the switch, and therefore, when a packet arrives, the switch already knows how to deal with it. This mode has negligible delay and removes the limits on the number of flows per second that can be handled by the controller.

In general, the SDN controller fills the flow table with the maximum number of possible flows. In reactive mode, flow setup is performed when a packet arriving at the switch does not match any of the switch entries. Then the controller will decide how to process/handle that packet, and the instructions will be cached onto the switch. As a result, reactive flow-setup time is the sum of the processing time in the controller and the time for updating the switch as the flow changes. Therefore, flow initiation adds overhead that limits network scalability and introduces reactive flow-setup delay [59, 60].

In other words, a new flow setup requires a controller to agree on the flow of traffic, which means that every flow now needs to go through the controller, which in turn instantiates the flow on the switch [61, 62, 63]. However, a controller is an application running on a server OS over a 10 GB/sec link (with a latency of tens of milliseconds). It is in charge of controlling a switch which could be switching 1.2 TB/sec of traffic at an average latency of 1μs. Moreover, the switch may deal with 100K flows, with an average of 30K being dropped. Therefore, a controller may take tens of milliseconds to set up a flow, while the life of a flow transferring 10MB of data (a typical Web page) is 10 msec [64, 52].

The authors in [57] carried out various setup experiments to test the throughput and latency of various controllers. They varied the number of switches, number of threads, and controller workload. Based on these experiments and simulations, they concluded that adding more threads beyond the number of switches does not improve latency and that serving a number of switches larger than the number of available CPUs increases controller response time [60]. The experiments also showed that controller response time varies between 4 and 30 ms for different number of switches with 4 threads and $2^{12}$ requests on the fly. However, the experimental setup and assumptions described in [60] need to be verified in realistic network environments.

Dealing with 100K flows requires that the switch ASICs must have this kind of flow capability. Current ASICs do not have this capability, and therefore the flow table must be used as a cache [64]. In conclusion, flow setup rate is anemic at best on existing hardware [64], and therefore only a limited number of flows per second are possible. The big O notation $O(n)$ linear lookup for software tables cannot approach the $O(1)$ lookup of a hardware-accelerated TCAM in a switch, causing a drop in the packet-forwarding rate for large wildcard table sizes [62].

To overcome performance limitations, the key factors that affect flow-setup time should be considered. As mentioned in [46], these key factors are the processing and I/O performance of the controller. Early benchmarks suggested that controller performance can be increased considerably by well-known optimization techniques such as I/O batching [60]. Another viable solution to alleviate the performance challenge was proposed under the name Maestro [65, 66]. Maestro used two basic parameters; the "input batching threshold" (IBT), a tuneable threshold value that determines the stage for creating a flow-task process to handle the flow request, and the "pending raw-packet threshold" (PRT) that determines the allowable number of pending packets in the queue to be processed. Calibration of these parameters will identify suitable values that will decrease latency and maximize network throughput according to network state. As the values of PRT and IBT increase, throughput increases and delays decrease [65]. Optimization techniques should be used to find the optimal range for values of PRT and IBT.

Finally, the DevoFlow and McNettle architectures described previously can be considered as feasible solutions to reduce network latency. McNettle implementations have shown that its improvements result in a 50-fold reduction in controller latency [50].

*5.4 Controlling the Data Path between the ASIC and the CPU*

Although the control data path in a line-card ASIC is fast, the data path between the ASIC and the CPU is not used in the frequent operations of the traditional switch, and therefore it is considered as a slow path. The ProCurve 5406lz Ethernet switch has a bandwidth of 300 GB/sec, but the measured loopback bandwidth between the ASIC and the CPU is 35 MB/sec [57]. Note also that the slow-switch CPU limits the bandwidth between the switch and the controller. For instance, the bandwidth of the flow-setup payload has been measured between the 5406lz switch and the OpenFlow controller and seems to be 10 MB/sec [57]. However, the DIFANE [54] architecture leverages these limitations by distributing the OpenFlow wildcard rules among various switches to ensure that forwarding decisions occur in the data plane.

Controlling the data path between the ASIC and the CPU is not a traditional operation [61]. OpenFlow specifies three counters for each flow-table entry: the number of matches, the number of packet bytes in these matches, and the flow duration. Each counter is specified as 64 bits, and therefore this adds 192 bits (24 bytes) of extra storage per table entry [67]. OpenFlow counters and the logic to support them add significant ASIC complexity and area and place more burdens on the CPU [63, 67, 68]. If counters are implemented in the ASIC hardware, it may be very difficult to change their function as the SDN protocol evolves because this would require re-designing the ASIC or deploying new switch hardware [67]. Moreover, transferring the local counter from the ASIC to the controller can dramatically limit SDN performance.

In addition, adding SDN support to create ASICs means finding space for structures not typically found on an ASIC; the per-flow byte counters used by OpenFlow could be the largest such structures. In other words, the counters take space from the ASIC area, in full knowledge that this area in considered precious because designing an ASIC costs a lot of money and time. However, because the cost of switch ASICs depends on their area, there is an upper limit on the area of a cost-effective ASIC [67]. Because ASIC area is valuable, this places limits on the sizes of on-chip memory structures such as TCAMs to support flow-table entries and per-entry counters. However, any silicon area allocated to counters will not be available for look-up tables [67].

As is well known, switches have a CPU to manage the ASICs, but the bandwidth between the two is limited [67]. Therefore, storing the counters in the CPU and DRAM instead of in the ASIC would simplify the path from the counters to the controller and minimize the overhead on the controller to access these counters. Another feasible solution that could address the limitations discussed above was suggested in [67]. The authors proposed software-defined counters (SDCs) because implementing counters in software does not require re-designing the ASIC and can support more innovations. In the proposed SDC, the ASIC does not contain any counters, but it does generate event records that will be added to the buffer. Whenever a buffer block is full, the ASIC moves it to the CPU. The CPU extracts the records and updates its counters, which are stored on the attached DRAM. SDC proposes two system designs:

*i)* A SDC switch in which the counter is moved out of the ASIC and replaced by buffer blocks.
*ii)* A SDC switch in which the CPU is installed on the ASIC. Although the second design requires additional ASIC space, it minimizes the bandwidth between the data plane and the CPU.

*5.5 Use of Low-Level Interfaces between the Controller and the Network Device*

Although SDN simplifies network management by developing control applications with simple interfaces to determine high-level network policies, the underlying SDN framework needs to translate these policies into low-level switch configurations [69]. The controllers available today provide a programming interface that supports a low-level, imperative, and event-driven model. The interface reacts to network events such as packet arrivals and link status updates by installing and uninstalling individual low-level packet-processing rules, rule-by-rule and switch-by-switch [70]. In such a situation, programmers must constantly consider whether un-installing switch policies will affect other future events monitored by the controller. Also, they must coordinate multiple asynchronous events at the switches to perform even simple tasks.

In addition, this interface generates a time-absorption problem and requires detailed knowledge of the software module or hardware device that is performing the required services. Many researchers are developing various programming languages that enable the programmer to describe network behaviour using high-level abstractions, leaving the run-time system and compiler to take care of implementation details. [71] proposes FML, a high-level programming language consisting of operators that allow or deny flows while coordinating the flows through firewalls and maintaining QoS. However, it is an inflexible language because it cannot redirect or move flows as they are processed [72].

Finally, Flog, an event-driven logic programming language, was proposed in [70]. Introducing logic programming to SDN is useful for processing network statistics and incremental controller-state updates. The main feature that differentiates

Flog from other languages is its Ethernet learning switch. The learning process consists of monitoring, grouping, and storing the packets that arrive at a switch and then transferring this information to a learning database. Afterwards, the policy generator creates low-level rules that flood all arriving packets, and then, based on the information learned, the policy creates a precise high-level forwarding rule.

*5.6 Controller Placement Problem*

The controller placement problem influences every aspect of a decoupled control plane, from flow-setup latencies to network reliability, to fault tolerance, and finally to performance metrics. For example, long-propagation-delay wide-area networks (WANs) limit availability and convergence time. This has practical implications for software design, affecting whether controllers can respond to events in real-time or whether they must push forwarding actions to forwarding elements in advance [73]. This problem includes controller placement with respect to the available network topology and the number of controllers needed. The user defines various metrics (latency, increase in the number of nodes, etc.) that control the placement of the controller in a network.

Random placement for a small $k$- value in the k-median problem, a clustering analysis algorithm, will result in an average latency between 1.4x and 1.7x larger than that of the optimal placement [73]. Finding the optimal controller placement is a hot SDN research topic, especially for wide-area SDN deployments because they require multiple controllers and their placement affects every metric in the network. Improving reliability is important because network failures cause disconnections between the control and forwarding planes and could disable some of the switches.

A reliability-aware controller placement problem has been proposed in [74]. The main objective of the problem can be understood using the following question: how to place a given number of controllers in a certain physical network such that the predefined objective function is optimized. The authors in [74] considered the reliability issue as a placement metric which is reflected by the percentage of valid control paths. They developed an optimization model that maximized the expected percentage of valid control paths. This percentage is affected by the location of the controller on one of the candidate nodes, the number of controller-to-controller adjacencies, the available number of controllers, and the reservation of the switches on the controller. Finally, a random placement algorithm and greedy algorithms have been suggested as heuristics to solve the reliable controller placement problem.

[75] states that any failure that disconnects the forwarding plane from the controller may lead to serious performance degradation. Based on this observation, the authors in [75] described a (path) resiliency (path protection)-aware controller placement problem. They considered connection resiliency between the controller and the switch as a placement metric which was reflected by the ability of the switches to protect their paths to the controller. The proposed heuristics aimed to maximize the possibility of fast failover based on resilience-aware controller placement and control-traffic routing in the network. These heuristics consisted of two algorithms for choosing the best controller location and maximizing the connection resiliency metric: the optimized placement algorithm and the approximation (greedy) placement algorithm.

Finally, the authors of [73] developed a latency-aware controller placement problem. Their objective was not to find the optimal solution for the latency-aware controller placement problem, but to provide an initial analysis for further study of the formulation of fundamental design problems. Therefore, the problem aimed to minimize the average propagation latency based on suitable controller placement. The minimization was based on an optimization model generated on the basis of the minimum $k$-median problem [73].

*5.7 Security*

Based on statistical studies carried out by IT organizations [76], 12% of respondents in IT business technologies stated that SDN has security challenges, and 31% of respondents in IT business technologies were undecided whether SDN is a less secure or a more secure network paradigm than others. Clearly, IT organizations believe that SDN may pose certain security challenges. According to the above studies, SDN security risks emerge from the absence of integration with existing security technologies and the inability to poke around every packet. Furthermore, improving the intelligence of the controller software may increase controller vulnerability to hackers and attack surfaces. If hackers access the controller, they will damage every aspect of the network, and it will be "game over" [76].

Increasing SDN security requires from the controller the ability to support the authentication and authorization classes of the network's administrators. In addition, leveraging the impact of security requires from the administrators the ability to use the same policies for traffic management to prevent access to SDN control traffic. Additional security-aware solutions are the implementation of an intelligent access control list (ACL) in the controller to filter packets and complete isolation between the tenants sharing the infrastructure. Finally, the controller should be able to alert the administrators in case of any sudden attack and to limit control communication during an attack [46].

SDN is a promising technology for computer networks and data-center networks, but it still lacks standardization policies. The current SDN architecture does not include standards for understanding topology, delay, or loss. Other features that are not available include loop detection and the ability to fix errors in a state. SDN does not support horizontal communications between network nodes to enable collaboration between devices [77].

As SDN gains in popularity, several researchers and enterprises have developed various SDN initiatives. They have proposed SDN prototypes, development tools, and languages for OpenFlow and SDN controllers and SDN cloud-computing networks [78]. Section VI covers some recent SDN implementations and tools.

## 6. RESEARCH INITIATIVES FOR SDN

SDN enables network owners and operators to build a simpler, customizable, programmable, and manageable network. According to the network research community, SDN will alter the future of networking and will import new innovations to the market [79]. With this in mind, a number of research initiatives have proposed SDN prototypes and applied them to DCN, wireless networking, software-defined radio, enterprises, and campus networks.

### 6.1 SDN Prototypes

The concept of SDN emerged in 2005, when the authors of [80] proposed a 4D approach to network control and management. Afterwards, a new network architecture, Ethane, which provides network control using centralized policies, was described in [81].

Ethane uses a centralized controller that holds network policies to control flow routing. It also uses Ethane switches which receive instructions from the controller to forward packets to their destinations. Policies are programmed using a flow-based security language based on DATALOG. Ethane was deployed in the Stanford computer science department to serve 300 hosts and in a small business to serve 30 hosts. Its deployment was an experiment to evaluate central network management, and it showed that a single controller could support 10,000 new flow requests per second for small network designs and that a distributed set of controllers could be deployed for large network topologies. Ethane has two limitations that prevent it from being implemented using current traditional network techniques. Initially, it requires knowledge about network users and nodes, and it demands control over routing at the flow level [82]. These limitations were addressed by NOX, a network operating-system framework.

Under NOX, applications can access the source and destination of each event, and routing modules can perform constrained routing computations. NOX makes it possible to build a scalable network with flexible control because it uses flows as its intermediate granularity [82].

### 6.2 Cloud Computing and Virtualization in SDN

Other recent studies [83] have developed an SDN-based controller framework, Meridian, for cloud-computing networks. Meridian provides a network services model that enables users to construct and manage a suitable logical topology for their cloud workloads. In addition, it allows virtual implementations on the underlying physical networks. Inspired by SDN, Meridian is composed of three logical layers: the network model and API layer, network orchestration, and interfaces to network devices. The first layer provides interaction with the network through declarative and query APIs; the declarative API creates the shape of the multi-virtual machine application, while the query API supports requests for topology views and network statistics. The orchestration layer provides services such as a global view of the data-center topology, routing algorithms, and scheduling network configuration and control functions. The lowest layer is responsible for creating virtual networks. In addition to the importance of Meridian in supporting a service-level model, it is considered as an initial prototype of SDN in the cloud. Researchers would like to explore the performance of Meridian in cases of sensitive workloads, the scalability of this framework to support large networks, and its ability to recover failed plans [83].

### 6.3 SDN Tools and Languages

Various tools and languages are used to monitor and implement SDN. Certain SDN initiatives have focussed on a forming platform, Onix, to implement SDN controllers as a distributed system for flexible network management [84]. Other studies have presented a network debugging tool, Veriflow [85], which is capable of discovering the faults in SDN application rules and hence preventing them from disrupting network performance. Additional initiatives [86] have developed a routing architecture, Routeflow, which is inspired by SDN concepts and provides interaction between commercial hardware performance and flexible open-source routing stacks. Hence, it opens the door to migration from traditional IP deployments to SDN.

In addition to recent studies that developed physical SDN prototypes, other researchers [62] have provided an efficient SDN innovation, Mininet. Mininet is a virtual emulator which provides an environment for prototyping any SDN idea. Whenever the prototype evaluation is acceptable, then it can be deployed in research networks and for general use [62]. However, Mininet's services are hindered by certain limitations: poor performance at high loads and its lightweight virtualization approach.

Research has also been directed toward developing control support for SDN and describing new language approaches to program OpenFlow networks.

[72] proposes a design for Frenetic, a high-level language for programming OpenFlow architectures. Frenetic consists of a query language based on SQL syntax, a stream-processing language, and a specification language for packet forwarding. With the combination of these three languages, Frenetic simplifies the programmer's task by enabling him/her to produce forwarding policies as high-level abstractions.

It addresses some of OpenFlow's shortcomings which are due to the lack of consistency between installing a rule in the switches and allowing other packets to be processed, in addition to the lack of synchronization between the packet arrival time and the rule installation time. It consists of two abstraction levels, the source-level operators that deal with network traffic, and the run-time system responsible for installing rules into switches.

In addition to the Frenetic language that can program OpenFlow networks, a number of other OpenFlow programming languages have been proposed, such as Procera [87, 88] and Nettle [58]. These languages are based on functional reactive programming, facilitate network management, and support event-driven networks.

*6.4 SDN Vendors*

[89] describes the Floodlight controller platform. It is an enterprise-class, Apache-licensed, Java-based OpenFlow controller that supports OpenStack orchestration and virtual and physical switches and manages OpenFlow and non-OpenFlow networks. In addition, NEC has designed a network virtualization architecture encapsulated as NEC ProgrammableFlow. The ProgrammableFlow technology provides management of their networking fabric. NEC has created custom physical switches, PF5240 and PF5820, to facilitate the ProgrammableFlow network architecture. The ProgrammableFlow controller can control any ProgrammableFlow or OpenFlow switch in a virtual network [90]. [91] provides an option list of existing OpenFlow controllers (NOX, Beacon [92], Helios, etc...) and switches (software and hardware options such as Open vSwitch, Pronto, etc…) to design SDN prototypes.

Besides these initiatives, researchers and enterprises have designed virtualization platforms for SDN [93]. NICIRA has created a complete SDN solution: the Network Virtualization Platform (NVP). It can be injected over existing network infrastructure or designed into emerging network fabrics. The NVP system works in collaboration with Open vSwitches that are configured in the hypervisor or used as gateways to legacy VLANs. Network virtualization is tasked to the Controller Cluster. The cluster is an array of control structures running on servers separate from the network infrastructure. Control is separated not only from network devices, but also from the network itself. Each cluster is capable of controlling thousands of Open vSwitch devices. The NVP architecture combines control and switching abstractions to provide a versatile network solution [94].

Finally, IT organizations and enterprises are focusing on applying SDN not only to data-center networks (LANs), but also to wireless local-area networks (WLANs) and wide-area networks (WANs), where OpenFlow will function as an overlay over L2 and L3 virtual private networks (VPN). HP has announced that an SDN-centralized controller can minimize the cost and complexity of implementing WAN optimization schemes. A prototype of SDN, Odin, was described in [95] and was intended to enable network operators to deploy WLAN services as network applications. Odin consists of a master, agents, and applications. The master runs as an application on the OpenFlow controller, controls the agents, and updates the forwarding table of access points (APs) and switches, and the agents run on the APs and collect information about the clients.

## 7. CONCLUSIONS

SDN aims to simplify network architecture by centralizing the control-plane intelligence of L2 switching and L3 routing equipment. It also markets network hardware as a product service and forms the basis of network virtualization. The generalized SDN architecture consists of the SDN controller and SDN-compatible switches. Because SDN makes it possible to build programmable and agile networks, academic researchers and network engineers are exploiting its flexibility and programmability to generate strategies that simplify the management of data-center LANs and WANs and make them more secure. Besides, SDN supports NaaS, the new Internet-based model that acts as a link between cloud computing and SDN. While SDN manages forwarding decisions and network administration, NaaS will provide packet-processing applications for cloud tenants. In addition, researchers are proposing various SDN prototypes that will serve DCNs, wireless networks, enterprises, and campus networks. Despite all the promising opportunities that accompany SDN, it encounters certain technical challenges that might hinder its functionality in cloud computing and enterprises. Therefore, IT organizations and network enterprises should be aware of these challenges and explore the functionality of the SDN architecture to counter these criticisms.

Communications Magazine, IEEE , vol.51, no.11, pp.46,52, November 2013.

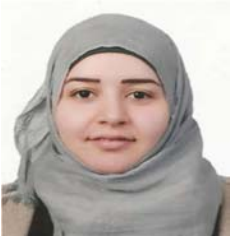

**MANAR JAMMAL** received her M.S. degree in electrical and electronics engineering in 2012 from Lebanese University, Beirut, Lebanon in cooperation with University of Technology of Compiègne. She is currently working towards the Ph.D. degree in cloud computing and virtualization technology at Western Ontario University. Her research interests include cloud computing, virtualization, software defined network and virtual machine migrations.

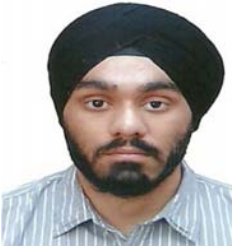

**TARANPREET SINGH** received his Masters in engineering (M.Eng) degree in Communications and Data Networking from the University of Western Ontario, London, Canada in September 2013. He has worked as a Consultant with Accenture Technology Services and holds special interest in the Cisco Networking domain. His research interests include Software Defined Networking, Network Function Virtualization and Network Security.

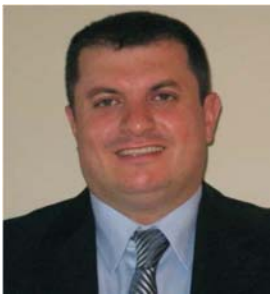

**Abdallah Shami** received the B.E. degree in Electrical and Computer Engineering from the Lebanese University, Beirut, Lebanon in 1997, and the Ph.D. Degree in Electrical Engineering from the Graduate School and University Center, City University of New York, New York, NY in September 2002. In September 2002, he joined the Department of Electrical Engineering at Lakehead University, Thunder Bay, ON, Canada as an Assistant Professor. Since July 2004, he has been with Western University, Canada where he is currently an Associate Professor in the Department of Electrical and Computer Engineering. His current research interests are in the area of network optimization, cloud computing, and wireless networks. Dr. Shami is an Editor for IEEE Communications Tutorials and Survey and has served on the editorial board of IEEE Communications Letters (2008-2013). Dr. Shami has chaired key symposia for IEEE GLOBECOM, IEEE ICC, IEEE ICNC, and ICCIT. Dr. Shami is a Senior Member of IEEE.

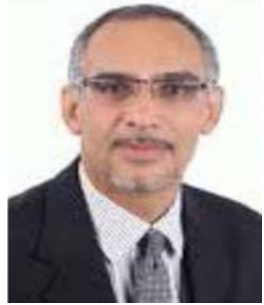

**Rasool Asal** received his PhD degree in physics from the University of Leicester (Leicester, UK). He then moved to the University of Southampton to take up a post-doctoral position within the Electronic and Computer Science Department. Dr. Rasool is a Chief Researcher at Etisalat BT Innovation Center (EBTIC) leading EBTIC research and innovation activities in the area of Cloud Computing. For the past fifteen years, he has been working with British Telecommunications Group at Adastral Park (Ipswich, U.K), designing and developing a considerable volume of high-performance enterprise applications, mostly in the area of telecommunications. Dr. Rasool is a speaker at many international conferences and events, most recently at the IEEE 8th International World Congress on Services (Cloud & Services 2012), Hawaii, USA. He has edited two books and published research papers in leading international conferences and journals. He is currently acting as Senior Guest Editor for Journal of Mobile Multimedia Special Issue on Cloud Computing Operation. His current interest focuses primarily on the Cloud Technologies, Cloud Security Architectures and the design of wide-area distributed cloud compliance enterprise systems that scale to millions of users.

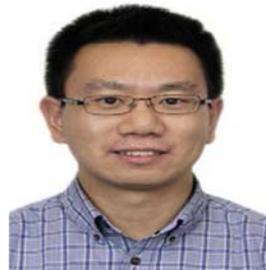

**Yiming Li** received a B.Eng. in Electrical Engineering from Western University, London, Ontario, Canada. Yiming is an Assistant Product Manager at StarTech.com. He is Responsible for product planning and product development. His research interests are in the areas of cloud computing, software defined networking and network virtualization.

24